\documentclass[aps,floatfix,pra,twocolumn,superscriptaddress]{revtex4-1}

\usepackage{amsmath}
\usepackage{amsfonts}
\usepackage{amssymb}
\usepackage{graphicx}
\usepackage{braket}

\begin{document}

\title{Generalized swap networks for near-term quantum computing}

\author{Bryan O'Gorman}
\affiliation{Department of Electrical Engineering and Computer Sciences, 
             University of California, Berkeley}
\affiliation{Department of Chemistry and Berkeley Quantum Information and Computation Center, 
             University of California, Berkeley}
\affiliation{Quantum Artificial Intelligence Laboratory, 
             NASA Ames Research Center, Moffett Field, CA}
\author{William J. Huggins}
\affiliation{Department of Chemistry and Berkeley Quantum Information and Computation Center, 
             University of California, Berkeley}
\author{Eleanor G. Rieffel}
\affiliation{Quantum Artificial Intelligence Laboratory, 
             NASA Ames Research Center, Moffett Field, CA}
\author{K. Birgitta Whaley}
\affiliation{Department of Chemistry and Berkeley Quantum Information and Computation Center, 
             University of California, Berkeley}

\date{\today}

\begin{abstract}
The practical use of many types of near-term quantum computers requires accounting for their limited connectivity.
One way of overcoming limited connectivity is to insert swaps in the circuit so that logical operations can be performed on physically adjacent qubits,
which we refer to as solving the ``routing via matchings'' problem.
We address the routing problem for families of quantum circuits defined by a hypergraph wherein each hyperedge corresponds to a potential gate.
Our main result is that any unordered set of $k$-qubit gates on distinct $k$-qubit subsets of $n$ logical qubits can be ordered and parallelized in $O(n^{k-1})$ depth using a linear arrangement of $n$ physical qubits;
the construction is completely general and achieves optimal scaling in the case where gates acting on all $\binom{n}{k}$ sets of $k$ qubits are desired. 
We highlight two classes of problems for which our method is particularly useful.
First, it applies to sets of mutually commuting gates, as in the (diagonal) phase separators of Quantum Alternating Operator Ansatz (Quantum Approximate Optimization Algorithm) circuits.
For example, a single level of a QAOA circuit for Maximum Cut can be implemented in linear depth, and a single level for $3$-SAT in quadratic depth.
Second, it applies to sets of gates that do not commute but for which compilation efficiency is the dominant criterion in their ordering.
In particular, it can be adapted to Trotterized time-evolution of fermionic Hamiltonians under the Jordan-Wigner transformation, and also to 
non-standard mixers in QAOA.\@
Using our method, a single Trotter step of the electronic structure Hamiltonian in an arbitrary basis of $n$ orbitals can be done in $O(n^3)$ depth while
a Trotter step of the unitary coupled cluster singles and doubles method can be implemented in $O(n^2 \eta)$ depth, where $\eta$ is the number of electrons.
\end{abstract}

\maketitle

\section{Introduction}

The state of experimental quantum computing is rapidly advancing towards ``quantum supremacy''~\cite{boixo2018characterizing}, i.e., the point at which quantum computers will be able to perform certain specialized tasks that are infeasible for even the largest classical supercomputers.
Beyond this technical milestone, however, lies another: \emph{useful} quantum supremacy, in which quantum computers can solve problems whose answers are of interest independently of how they were achieved.
The combination of efficient quantum algorithms~\cite{shor1999polynomial-time,grover1996fast} and scalable error correction~\cite{fowler2009high-threshold} makes such progress likely in the long term, barring fundamental surprises.
In the near term, we have so-called Noisy Intermediate-Scale Quantum (NISQ) devices~\cite{preskill2018quantum}, capable perhaps of outperforming classical devices on certain problems, but with extremely constrained resources. 
Many types of such devices (e.g., superconducting quantum processors) will have limited connectivity.
For the most part, existing quantum algorithms assume an abstract device with arbitrary connectivity, i.e., the ability to do a two-qubit gate between any pair of qubits.
In theory, this suffices given that circuits can be compiled to any concrete family of devices with polynomial overhead in qubits and gates~\cite{brierley2017efficient}.
In practice, polynomial overheads matter and can be the crucial difference between being feasible and infeasible on NISQ devices.

The overall goal of compilation within the quantum circuit model is to take a quantum algorithm and implement it (maybe approximately) on a concrete piece of quantum computing hardware.
There are many approaches to this, but perhaps the most straightforward is to transform the desired quantum circuit into an executable one in two steps: 1)~\emph{decomposition} of the constituent gates into (maybe approximately) equivalent sub-circuits consisting of ``native'' gates, and 2)~what we call \emph{routing via matchings} of the circuit~\cite{childs2019circuit}. 
Our focus here is on the routing problem, in which the logical qubits are dynamically assigned to physical qubits in a way that allows the desired logical gates to be implemented while respecting the restricted connectivity of the actual hardware. 
In general, it may be necessary to use swap gates to change this assignment of logical qubits to physical qubits throughout the execution of the circuit.

In the past several years, there has been a blossoming of tools for addressing variants of this routing problem, which are variously called ``quantum circuit placement'', ``qubit mapping'', ``qubit allocation'', or ``quantum circuit compilation'' (though the latter term generally encompasses much more).
Prior work, however, has taken one of two approaches. 
First, of theoretical interest, is to show how \emph{any} quantum circuit can be converted ``efficiently'' (i.e., with polynomial overhead) into one in which gates act only locally in some hardware graph~\cite{brierley2017efficient,hirata2009efficient,maslov2007quantum}.
The second is an instance-specific approach, in which the problem is solved anew for each logical circuit~\cite{bhattacharjee2017depthoptimal,siraichi2018qubit,li2018tackling,lye2015determining,venturelli2018compiling,booth2018comparing,saeedi2011synthesis,wille2016look-ahead,lin2015paqcs,zulehner2018efficient,herbert2018using}.
We propose and instantiate a new instance-independent approach, in which the routing is done for a family of instances, with little-to-no compilation necessary for each instance; the per-instance compilation time is therefore effectively amortized to nil. 
This approach, which finds solutions for families of instances, interpolates between the two approaches above and seeks to balance the time to solution and the quality of solution.
The family-specific routing can be found either algorithmically or, as is done here, manually.
Algorithms useful in the instance-independent approach, where quality of solution is prioritized over time to solution, may (but not necessarily) differ significantly from those useful in the instance-specific approach, wherein the prioritization is reversed. On the other hand, for many problem families, there is an instance with maximal structure on which instance-specific algorithms can be run, thus obtaining compilations that can be used for the whole family. In general, these instance-specific approaches work best on sparser cases, and on dense instances will return inferior compilations to the ones given here.  

In many quantum algorithms for quantum chemistry it is the case that all circuits of a given size for a particular problem have the same structure with respect to a partial ordering of the operations, and the only instance-specific aspect is the parameters (e.g.\ rotation angles) of the gates.
Furthermore, the implementation of these gates on hardware often has the same properties (e.g.\ fidelity and duration) regardless of the parameters.  
In such cases, the instance of the compilation problem is effectively independent of the instance of the application problem.
Compare this with implementing QAOA on hardware in which gate durations are independent of their parameters, in which case the routing problem for a given problem instance is the same regardless of the variational parameters, but differs significantly for different problem instances. In cases in which gate durations vary, an upper bound on (or average over) the range of durations can be used to obtain instance-independent compilations.
Thus, the distinction between instance-specific and instance-independent approaches is somewhat subjective and contextual, but we merely aim to emphasize that there is an under-explored regime in the trade-off between quality of solution and computation time in approaches to the quantum circuit routing problem. 

An alternative approach for variational algorithms in general is to obviate the compilation problem by using an ansatz that is based on the connectivity of the target hardware~\cite{farhi2017quantum,kandala2017hardware} and less so on the target application.
By efficiently compiling application-specific circuits to constrained hardware, our methods combine the efficiency of this approach with respect to physical resources with the advantages of an application-specific ansatz (e.g.\ fewer variational parameters).

A method was recently proposed for implementing a Trotter step of a fermionic Hamiltonian containing \(\binom{n}{2}\) terms, where \(n\) is the number of orbitals, using a circuit of depth \(n\) with only linear connectivity~\cite{kivlichan2018quantum,jiang2018quantum}.
Using fermionic swap gates~\cite{corboz2009fermionic}, Kivlichan et al.\ were able to change the mapping between fermionic modes and physical qubits while preserving anti-symmetry~\cite{kivlichan2018quantum}. 
By constructing a network of these gates such that, at some point in the circuit,
each pair of orbitals is assigned to some pair of neighboring qubits, they were able to guarantee that they could implement each of the terms in the Trotter decomposition of the Hamiltonian by acting only locally on said pair of qubits, and that they could implement \(n/2\) such terms in parallel.
In this work, we generalize their approach and describe a way to construct networks of fermionic swap gates acting on \(n\) qubits such that each \(k\)-tuple of fermionic modes is mapped to adjacent physical qubits at some point during the circuit. 
The circuits that we construct have an asymptotically optimal depth of \(O(n^{k-1})\) while only assuming linear connectivity.

For fermionic Hamiltonians, Motta et al.\ take a different approach that exploits the fact that many Hamiltonians of practical interest are low-rank~\cite{motta2018low-rank}.
For unitary coupled cluster and full-rank generic chemical Hamiltonians, their methods achieve the same scaling as ours, as summarized in Table~\ref{tbl:main-results}.
Our methods provide an alternative Trotter order, whose relative value will need to be studied empirically.
For a Trotter step of the Hamiltonian for real molecular systems, empirical data indicate that their low-rank methods can achieve $O(n^2)$ depth. 

The question of how to optimally implement a collection of $k$-qubit operators is not confined to the simulation of fermionic quantum systems. 
Another promising use is in the application of the Quantum Alternating Operator Ansatz (Quantum Approximate Optimization Algorithm, or QAOA) to Constraint Satisfaction Problems (CSPs) over Boolean domains.
This approach was taken for the Maximum Cut problem using existing linear swap networks~\cite{crooks2018performance}; our methods can address $k$-CSP for any $k$ in $O(n^{k-1})$ depth.

Our main contributions are:
\begin{itemize}
\item Formalizing a variant of the quantum circuit routing problem in a way that abstracts away details of particular devices and focuses on their geometry, which is shared by a wide class of devices;
\item Making explicit and general the equivalence between swap networks that change the mapping of logical qubits to physical qubits and those that change the mapping between fermionic modes and physical qubits;
\item Explicit constructions for several important classes of problems, as summarized in Table~\ref{tbl:main-results}, using modular primitives that can be applied to new problems; and
\item Providing tools for lower bounding the depth of solutions to the routing problem, in particular by connecting it with prior work on acquaintance time and graph minors.
\end{itemize}

This paper is organized as follows.
In Section~\ref{sec:model}, we more formally describe the quantum circuit routing problem and our approach thereto.
In Section~\ref{sec:swap-networks}, we introduce generalized swap networks that will be used in the constructions of later sections.
In Section~\ref{sec:problem-families}, we introduce some specific quantum simulation tasks related to fermionic Hamiltonians, as well as the Quantum Alternating Operator Ansatz (QAOA), which yield families of circuits that can be routed using our methods.
In Section~\ref{sec:complete-hypergraphs}, we present our main result, showing how to achieve optimal scaling when routing (with an arbitrary ordering) circuits consisting of a $k$-qubit gate for each possible set of $k$ qubits.
In Section~\ref{sec:unitary-coupled-cluster} we describe families of instances arising from the Unitary Coupled Cluster method and how to efficiently route them.
In Section~\ref{sec:conclusion}, we conclude.
A reader familiar with either QAOA or quantum simulation of fermionic Hamiltonians and interested in quickly learning some useful techniques may do so in sections~\ref{sec:swap-networks},~\ref{sec:complete-hypergraphs}, and~\ref{sec:unitary-coupled-cluster}.

In Appendix~\ref{sec:quantum-annealing}, we discuss the instance-independent approach in the context of quantum annealing.
In Appendix~\ref{sec:lower-bounds}, we show how to lower bound the depth of a solution to the circuit routing problem.

\begin{table}
\begin{tabular}{lcccc}
\hline \hline 
Instance family: & $k$-CSP& UCCGSD & UCCSD & UpCCGSD\\
Depth: & $\Theta(n^{k-1})$ &  $\Theta(n^3)$ & $\Theta(\eta n^2)$ & $\Theta(n)$\\
\hline \hline 
\end{tabular}
\caption{
Main results.
$k$-CSP indicates depth for a single round of QAOA.\@
Remaining columns indicate depth for a single Trotter step of the coupled cluster operator or of a similarly structured variational ansatz. 
All are optimal up to constant prefactors for arbitrary connectivity.
See Section~\ref{sec:complete-hypergraphs} for details regarding $k$-CSP and Section~\ref{sec:unitary-coupled-cluster} regarding Unitary Coupled Cluster.
}\label{tbl:main-results}
\end{table}

\section{Model}\label{sec:model}

\begin{figure}
 \includegraphics[width=\columnwidth]{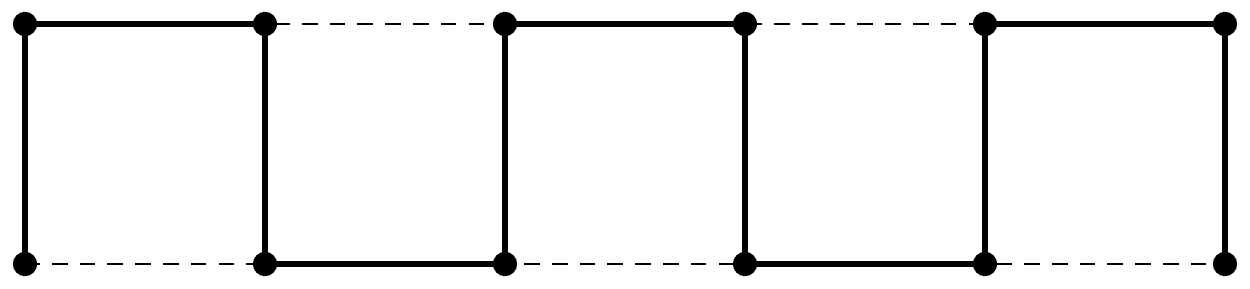}
 \caption{\label{fig:castellation}
 ``Castellated'' mapping of the Jordan-Wigner string into a $2\times (n / 2)$ square lattice.
 While all swapping is done along the Jordan-Wigner string, mapping the string to the lattice in this way allows for potentially more efficient decomposition of the $4$-qubit gates by making available a fourth adjacency.
 }
\end{figure}

We consider hardware consisting of a line of $n$ qubits and suppose that we are able to implement any $k$-qubit  gate in time $\tau_k$ on any set of $k$ qubits that are adjacent.
This is an abstraction of the more physical model in which only $1$- and $2$-qubit gates can be directly implemented; $\tau_k$ for $k > 2$ is thus some linear combination of $\tau_1$ and $\tau_2$ that indicates an upper bound on the cost of compiling any $k$-qubit gate.
When considering a specific piece of hardware, this model is relatively coarse; different gates on different sets of physical qubits may require vastly different times to implement.
However, this level of abstraction allows for significant generality without too great a loss of precision.
Accordingly, for a specific piece of hardware, our constructions should be considered as a starting point, with low-level optimizations likely to improve the constant factors significantly.
For example, the line of qubits on which the swap networks are defined can be embedded in a ``castellated'' manner in a $2 \times (n / 2)$ lattice, as shown in Figure~\ref{fig:castellation}.
The availability of the additional qubit adjacency can enable more efficient decomposition of higher-locality gates.

The problem we would like to solve is as follows:
given a set of $k$-qubit gates $G$ on $n$ qubits, implement them in some order on the hardware described above.
In particular, we focus on the swap-network paradigm.
That is, we start with an initial assignment of logical qubits to physical qubits and insert a sequence of $2$-qubit swap gates to move the logical qubits around so that for every gate in $G$ the logical qubits on which it acts are physically adjacent at some point in the process.
As discussed in Sections~\ref{subsec:fermionic-hamiltonians} and~\ref{sec:unitary-coupled-cluster}, the routing problem thus defined is equivalent to the problem of using fermionic swap gates to change the ordering of a Jordan-Wigner string to enable the implementation of gates locally.
Without loss of generality, we assume that there is at most one gate in $G$ acting on any set of qubits, and that for any gate $g \in G$ acting on a set of qubits $S$ there is no other gate $g' \in G$ acting on a subset of qubits $S' \subset S$.
This is a convenient abstraction, rather than a restriction.
An instance of the routing problem is thus specified as a hypergraph, with vertices corresponding to logical qubits and hyperedges corresponding to logical gates. We focus on $k$-complete hypergraphs, ones in which for every subset of $k$ vertices, there is an edge connecting them; $|G| = \binom{n}{k}$.
Results for complete hypergraphs give worst case bounds for the general problem. 
A more general variant is the more typically considered problem in which one wants to enforce a temporal partial ordering on the logical gates.

In general, near-term hardware will have greater connectivity than a line; 
nevertheless, it will likely contain a line as a subset, so that our constructions give a baseline.
Even with greater connectivity, our scaling is optimal when the number of gates is $\Omega(n^k)$.
Let $m = |G|$ be the number of gates, $\nu$ the number of physical qubits, and $n$ the number of logical qubits.
At most $\nu$ gates can be implemented at a time, so the circuit depth must be at least $m / \nu$.
For $m = \Omega(n^k)$ and $\nu = O(n)$, this implies a minimal depth of $\Omega(n^{k-1})$, which our construction provides.
Because our focus is on resource-constrained near-term hardware, we shall assume that the number of physical qubits is equal to the number of logical qubits.

\section{Swap networks}\label{sec:swap-networks}

\begin{figure*}
\includegraphics[width=\textwidth]{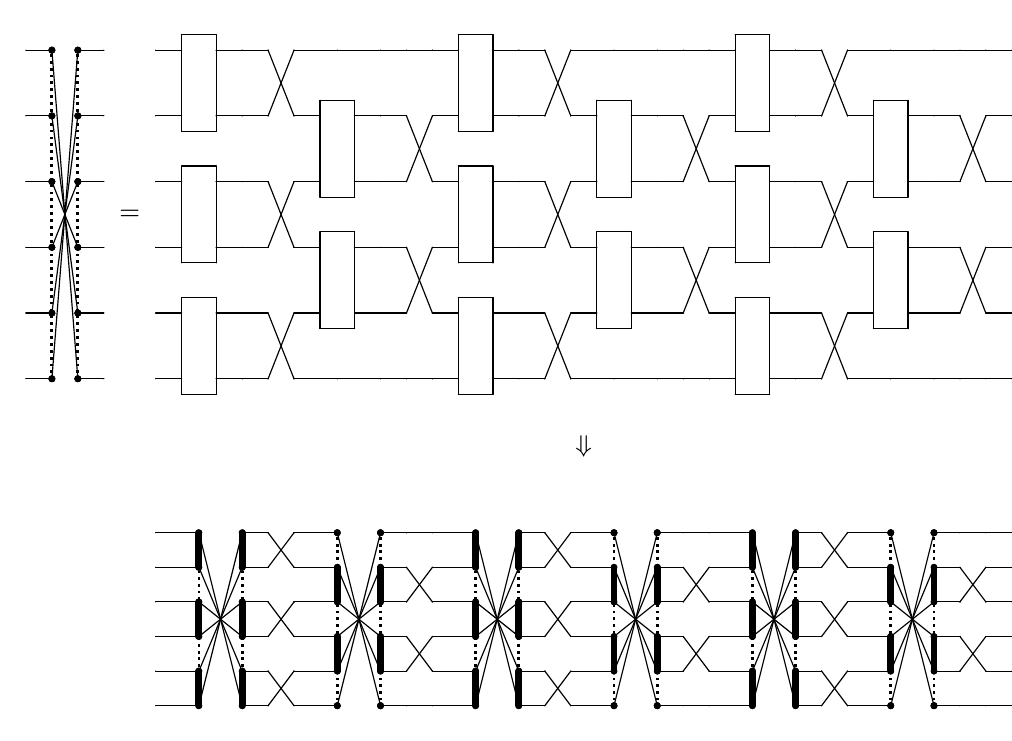}
\caption{\label{fig:3-local}
Top:
Notation (left) and decomposition (right) for the canonical 2-complete linear (2-CCL) swap network for acquainting all pairs of qubits, annotated with empty boxes showing acquaintance opportunities.
The circuit has depth $n$, and contains $n(n-1)/2$ swap gates.
Bottom:
The canonical 3-complete linear (3-CCL) swap network that acquaints all triples of qubits.
It is formed by replacing each layer $i$ of acquaintance opportunities in the 2-CCL swap network with a $\mathcal P_i$-swap network;
the qubits to be acquainted in the former are the parts in the partition $\mathcal P_i$.
The $3$-local acquaintance opportunities are not shown.
Part of this swap network is shown in more detail in the top of Figure~\ref{fig:4-local}, with acquaintance opportunities indicated. 
(Each layer of $1$-swaps will be cancelled out by the final layer in the expansion of the preceding $2$-swap network; we include them here to make the recursive step clear.)
}
\end{figure*}

Henceforth, by ``swap gate'', we shall mean either the standard swap gate (when considering a mapping of logical qubits to physical qubits) or the fermionic swap gate (when considering a mapping of fermionic modes to physical qubits); for circuit routing, everything is exactly the same in both cases except for the ``interpretation''.
A \emph{swap network} is a circuit consisting entirely of swap gates. We define a \emph{2-complete linear swap network}, a notion we shall generalize shortly, to be a swap network in which all pairs of logical qubits are linearly adjacent at some point in the circuit and in which all swap gates act on linearly adjacent physical qubits. 
Such networks ensure that, in the linear architecture described in Sec~\ref{sec:model}, there is an opportunity to add, for each pair of logical qubits, a 2-qubit gate acting on those logical qubits (or fermionic modes as the case may be) at some point in the circuit. We call such opportunities \emph{acquaintance opportunities}.
They are not part of the swap network, but we shall often draw them as empty boxes in circuit diagrams to illustrate acquaintance properties of swap networks, as in Figure~\ref{fig:3-local}.
We shall say that a set of logical qubits that has at least one such acquaintance opportunity is ``acquainted'' by the network, or that the swap network ``acquaints'' those qubits.

Before generalizing this notion, we review the construction of Kivlichan et al.~\cite{kivlichan2018quantum}
for implementing a $2$-local gate on every pair of logical qubits in depth $n$, using $\binom{n}{2}$ swap gates.
The swap network underlying this construction is what we shall call the \emph{canonical} 2-complete linear swap network.
Let the physical qubits be labeled $1$ through $n$, and partition the pairs of adjacent qubits into two sets based on the parity of their larger index: 
even pairs $\left\{\{1, 2\}, \{3, 4\}, \ldots\right\}$ and
odd pairs $\left\{\{2, 3\}, \{4, 5\}, \ldots, \right\}$.
Note that the pairs in each partition are mutually disjoint.
We define the \emph{canonical 2-complete linear} (2-CCL) swap network as $n$ alternating layers of swaps on the even pairs and odd pairs, as illustrated in the top half of Figure~\ref{fig:3-local}. 
The overall effect of the 2-CCL swap network is to reverse the ordering of the logical qubits.
In doing so, it directly swaps every pair of logical qubits.
This construction has the attractive property that each acquaintance opportunity precedes a swap gate on the same two qubits, so any added gate that acts on a pair of logical qubits can be combined with the swap of those two qubits, with the result that in depth $n$ we can execute a $2$-qubit gate between every pair of logical qubits.

One direction for generalization is to $({\cal S},{\cal A})$-swap networks, where $\cal S$ is a subset of all pairs of qubits and $\cal A$ is an architecture, such as a 2D grid.
The set $\cal S$ captures the pairs of qubits to which we want to apply 2-qubit gates at a given stage in a circuit.
We shall not discuss this generalization further in this paper, other than to note that our results can be used to provide bounds for $({\cal S},{\cal A})$-swap networks.
Because in the present work we shall present only swap networks acting on a line, we shall often leave that aspect implicit in the terminology and refer simply, e.g., to a ``2-complete swap network''.
 
Instead, we are interested in generalizing to \emph{$k$-complete swap networks}, networks in which the elements of every set of $k$ logical qubits are adjacent at some point, so that a $k$-qubit gate (or set of $1$- and $2$-qubit gates making up the $k$-qubit gate) could be applied thereto.
To support the construction of $k$-complete swap networks in Sec.\ref{sec:complete-hypergraphs}, here
we introduce a generalization of a 2-complete swap network that swaps elements of a partition of qubits, rather than individual logical qubits: a \emph{complete $\mathcal P$-swap network}, where $\mathcal P$ is an ordered partition of the physical qubits such that each part contains only contiguous qubits, contains only swap operators that swap parts of the partition.
In this way, a complete $\mathcal P$-swap network has the property that every part in the partition is adjacent to every other part in the partition at some point in the network.

\begin{figure}
\includegraphics[width=0.9\columnwidth]{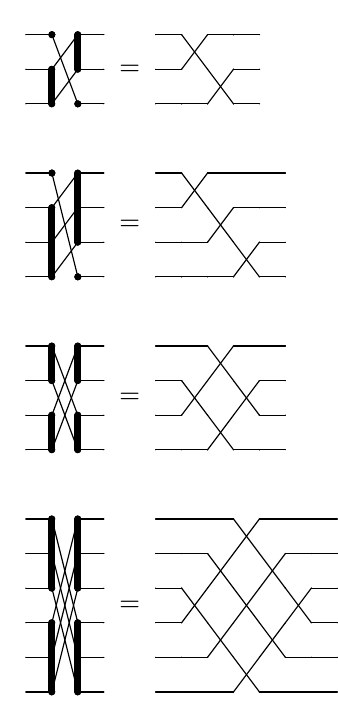}
\caption{\label{fig:k-swaps}
{\bf Generalized swap gates.} From top to bottom, notation and decompositions for $(1,2)$-, $(1, 3)$-, $(2, 2)$-, and $(3, 3)$-swap gates.
A $(k_1,k_2)$-swap gate can be implemented in depth $k_1+k_2-1$ using $k_1 k_2$ standard swap gates.
} 
\end{figure}

In constructing $\mathcal P$-swap networks, it will be useful to swap pairs of sets of qubits using what we call a \emph{$(k_1, k_2)$-swap gate}, or, more generally, a \emph{generalized swap gate}.
The $(k_1, k_2)$-swap gate swaps a set of $k_1$ logical qubits with a set of $k_2$ logical qubits, while preserving the ordering within each set, i.e., it permutes a sequence of logical qubits from  
$\left(i_1, \ldots, i_{k_1}, i_{k_1 + 1}, \ldots, i_{k_1 + k_2}\right)$ to 
$\left(i_{k_1 + 1}, \ldots, i_{k_1 + k_2}, i_1, \ldots, i_{k_1}\right)$.
Several examples of these generalized swap gates and their decompositions are shown in Figure~\ref{fig:k-swaps}.
In general, a $(k_1, k_2)$-swap gate can be decomposed using $k_1 \cdot k_2$ standard swap gates in depth $k_1 + k_2 - 1$.
We call a swap network a \emph{$k$-swap network} whenever it contains only $(k_1, k_2)$-swap gates for $k_1, k_2 \leq k$. 

The \emph{canonical $\mathcal P$-swap network} has the same structure as the 2-CCL swap network, except that instead of pairs of single qubits being swapped at a time, pairs of sets of qubits (i.e., the parts of the partition) are swapped.
In the canonical $\mathcal P$-swap network, each $(k_1, k_2)$-swap gate is preceded by a $(k_1 + k_2)$-local acquaintance opportunity. 
To make the overall effect of a complete $\mathcal P$-swap network be a complete reversal of the qubit mapping, we append to the end a $1$-swap network within each part.
This is unnecessary when considering a single swap network, but may be helpful when using the swap network as a primitive in a larger construction.
Note that this is is primarily for explanatory purposes, and in an actual implementation would likely be optimized away.
In the recursive strategy for $k$-local hypergraphs (discussed in Sec.~\ref{sec:complete-hypergraphs}), each generalized swap gate is preceded by some number of acquaintance opportunities and swap gates that ensure that each set of $k_1$ or $k_2$ qubits is acquainted with each one of the other set.

The 2-CCL swap network has the exact same structure as the optimal sorting network on a line~\cite{beals2013efficient}.
A sorting network is a fixed circuit consisting of ``comparators''.
Given an initial assignment of objects to the wires, each comparator compares the objects and swaps them if they are out of order.
This means that a subset of the 2-CCL swap network can be used to effect an arbitrary permutation of logical qubits in at most linear depth.

\begin{figure}
\includegraphics[width=\columnwidth]{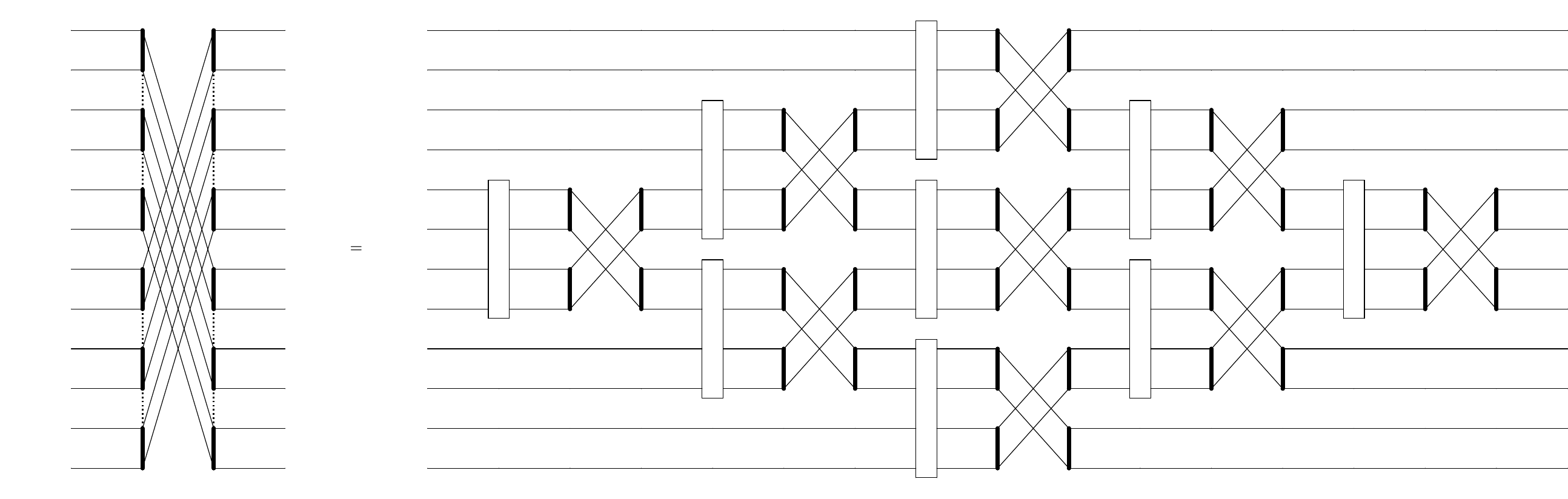}
\caption{\label{fig:bipartite-swap-network}
Notation (left) and decomposition (right) for a bipartite swap network corresponding to the bipartition 
$\left(
    \left(\left(1, 2\right), \left(3, 4\right), \left(5, 6\right) \right),
    \left(\left(7, 8\right), \left(9, 10\right), \left(11, 12\right) \right)
\right)$.
For each pair of qubits in the first half and each pair of qubits in the second half, their union is acquainted; the overall effect is the same as that of a generalized swap gate corresponding to the two halves.
Note the similarity of the notation to that for a complete swap network, except that the dotted lines connect only each part of the bipartition.
}
\end{figure}

The swap networks above acquaint all pairs of sets of qubits.
Another useful primitive is what we call a ``bipartite swap network''; again, this should be more precisely called a ``bipartite linear swap network'' to emphasize that it acts on a line, but we leave this implicit for concision.
Given a bipartition of sets of qubits, it acquaints all the unions of pairs of sets of qubits which can be formed by taking one set from the first part and the other set from the second part.
While the depth of a bipartite swap network is similar to that of a complete swap network, the gate count is approximately halved.
Figure~\ref{fig:bipartite-swap-network} shows an example bipartite swap network for the sets of qubits $\left((1, 2), (3, 4), \ldots, (11, 12)\right)$ with the first three in one part and the latter three in the second part.

Swap networks can be useful for measurement as well.
In many cases, the gates to be executed correspond one-to-one with the terms of a Hamiltonian to be measured.
Any swap network used to implement those gates thus yields a partition of the terms of the Hamiltonian into parts containing only gates acting on disjoint sets of qubits.
This partition can then be used to parallelize the measurements.
After an application of the swap network, the swap layers following the logical layer to be measured can be executed in reverse to return the mapping to one in which the terms of the Hamiltonian are mapped to adjacent sets of qubits, with appropriate optimizations made to account for the fact that many swap gates will likely cancel out once the logical gates are removed.
Alternatively, a simple sorting network can be used to achieve the same end.
For fermionic Hamiltonians, this approach can significantly reduce the number of measurements needed by reducing the locality of all measurement terms, in addition to the savings yielded by parallelization.

\section{Problem families}\label{sec:problem-families}

\begin{table}
\begin{tabular}{lcccc}
\hline \hline 
Application & QAOA & Quantum chemistry \\
\hline 
Iteration & 
    $\prod_I \exp\left(i \gamma c_I \prod_{i \in I} Z_i \right)$
    &
    $\prod_{p, q, r, s} \exp\left( - i t H_{p, q, r, s}\right)$ \\
Assignment & logical qubits & fermionic modes \\
Changed by & SWAP & FSWAP \\
\hline \hline 
\end{tabular}
\caption{The two problem families we consider. The table lists the iterated operator we compile, and the logical unit and gates used in the compilation.
}\label{tbl:swap-types}
\end{table}

In this section, we introduce two families of quantum circuits that come from quite different application domains but whose compilation can be addressed using essentially the same tools;
the analogy is summarized in Table~\ref{tbl:swap-types}.
Both cases involve repeated application of a circuit of a particular form such that for each iteration the compilation instance is the same.
We provide constructions for a single iteration; these can be repeated sequentially for the full circuit.
Solving the compilation instance for the full circuit all at once may provide a better solution, but likely at the cost of it being much harder to find.

\subsection{Fermionic Hamiltonians}\label{subsec:fermionic-hamiltonians}
The general form of the electronic structure Hamiltonian in second quantization is 
\begin{equation}\label{eq:fermion-ham}
H = 
\sum_{p, q} c_{p, q} a^{\dagger}_p a_q + 
\sum_{p, q, r, s} c_{p, q, r s} a^{\dagger}_p a^{\dagger}_q a_r a_s,
\end{equation}
where $p,q, r, s$ label single-electron orbitals, 
$c_{p, q}$ and $c_{p, q, r, s}$ are real coefficients, and
$a^\dagger_p$ is the creation operator for the $p$th orbital.
A common subroutine of quantum simulation algorithms is the Trotterization of time evolution under such a fermionic Hamiltonian~\cite{aspuru2005simulated}:
\begin{equation}
e^{-i H t } 
\approx
\prod_{l=1}^{t / \delta t}  \left(\prod_{p, q, r, s} e^{-i H_{p,q, r,s} \delta t} \right),
\label{eq:trotterized_evolution}
\end{equation}
where $H_{p,q,r,s}$ is the part of the Hamiltonian that acts exclusively on modes $p$,$q$,$r$,$s$. 
(For simplicity we absorb the terms acting on two fermionic modes into the terms acting on four.)
One approach to mapping the fermionic operators \(e^{-iH_{p,q,r,s}}\) into operators acting on the qubit Hilbert space is to employ the
Jordan-Wigner transformation~\cite{ortiz2001quantum},

\begin{equation}
    a_p = -\prod_{i=1}^{p-1}\sigma_i^{(z)} \cdot \sigma_p^{(-)}.
\end{equation}
After performing the Jordan-Wigner transformation on Equation~\ref{eq:trotterized_evolution}, many of the resulting
operators will act non-trivially on \(\Theta(n)\) qubits, resulting in a naive gate depth of \(\Theta(n^5)\) for the implementation of Equation~\ref{eq:trotterized_evolution}, assuming there are \(\Theta(n^4)\) terms in the Hamiltonian.
As we shall see, by reordering the fermionic modes (thereby changing the
Jordan-Wigner ordering), this overhead from the non-locality of the
Jordan-Wigner transformation is addressed automatically in our scheme for parallelization.
For this reason, our constructions provide significant advantage even when connectivity is not a constraint, including in the error-corrected regime.
As a result, at least with respect to scaling, we avoid the need for more sophisticated alternatives to the Jordan-Wigner transformation, such as those developed by Bravyi and Kitaev~\cite{bravyi2002fermionic} and others~\cite{bravyi2017tapering}.

A related approach, employed by a variety of works proposing the study of quantum chemistry using a near-term device, is the use of a quantum circuit to prepare and measure the unitary coupled cluster ansatz~\cite{peruzzo2014variational,mcclean2016theory, romero2018strategies,lee2018generalized}.
Under the typical choice to include only single and double excitations in the cluster operator, this wave function is given by
\begin{equation}
    |\psi\rangle = e^{T - T^\dagger}|\phi_0\rangle,
    \label{eq:unitary_coupled_cluster}
\end{equation}
where the cluster operator \(T\) has a form similar to \(H\) in Equation~\ref{eq:fermion-ham}.
Usually, it contains only excitations from the $\eta$ ``occupied'' orbitals which contain an electron in the reference state \(|\phi_0\rangle\) to the \(n - \eta\) ``virtual'' orbitals, and the coefficients are determined variationally. 
We refer to this case as UCCSD, and the case where all $2$-electron excitations are included as UCCGSD.\@
The exact exponential of Equation~\ref{eq:unitary_coupled_cluster} is typically approximated by a Trotter expansion and (assuming \(n \gg \eta\)), the \(\Theta(n)\) overhead from the non-locality of the Jordan-Wigner strings discussed above would lead to a circuit depth of \(\Theta(n^3\eta^2)\) for a single Trotter step. 

We show how depths of \(O(n^3)\) and \(O(n^2\eta)\) can be achieved for a Trotter step of the time evolution under a fermionic Hamiltonian (or the similarly structured UCCGSD) and the UCCSD ansatz, respectively.
These scalings match the asymptotic results of Ref.~\cite{hastings2015improving} while also respecting the spatial locality of the available gates and requiring no additional ancilla qubits.

\subsection{QAOA}
As originally proposed~\cite{farhi2014quantum,farhi2016quantum}, QAOA is a method for minimizing the expectation value of a diagonal Hamiltonian
\begin{equation}
H_f = \sum_{I \subset [n]} c_I \prod_{i \in I} Z_i
\end{equation}
corresponding to a classical function $f: {\{\pm 1\}}^n \to \mathbb R$ whose multilinear form is
\begin{equation}
f(\mathbf s) = \sum_{I \subset [n]} c_I \prod_{i \in I} s_i.
\end{equation}
The minimization is done variationally over states of the form
\begin{equation}
\ket{\boldsymbol \beta, \boldsymbol \gamma} =
e^{i \beta_p M} e^{i \gamma_p H_f}
\cdots
e^{i \beta_1 M} e^{i \gamma_1 H_f}
{\ket{+}}^{\otimes n},
\end{equation}
which consists of $p$ alternating applications of the ``phase separator'' $e^{i\gamma H_f}$ and the ``mixer'' $M = \sum_{i = 1}^n X_i$.
The phase separator can be written as the product of gates corresponding to terms in the Hamiltonian,
\begin{equation}
e^{i \gamma H_f} = 
\prod_{I} e^{i \gamma c_I \prod_{i \in I} Z_i}.
\end{equation}
Note that the gates are diagonal and so their order does not matter.
The locality of the gates corresponds directly to the locality of the terms in the Hamiltonian, $\max|I|$.
QAOA applied to $k$-CSP, in which each term acts on at most $k$ variables, thus requires $k$-qubit gates.

Hadfield et al.~\cite{Hadfield17_QAOAGen} generalized QAOA to the Quantum Alternating Operator Ansatz, employing a wider variety of mixers, many of which involve $k$-qubit gates.
While these gates, in general, do not commute, it is an open question how the order of the gates affects the efficacy of the mixing.
In NISQ devices with limited depth, the depth in which different mixers can be implemented plays a key role in their usefulness.
The techniques here can be applied to these alternative mixers, with different orderings giving different mixers within the same family, and the resulting compilation a key step in determining the most effective mixing strategy.

\section{Complete hypergraphs}\label{sec:complete-hypergraphs}
\subsection{Cubic interactions}

Now, suppose we want to implement a $3$-qubit gate between every triple of logical qubits. We call a swap network that achieves this goal a \emph{$3$-complete linear} (3-CCL) swap network.
We can do so in the following way.
First, we start with the 2-CCL swap network, as shown in the top half of Figure~\ref{fig:3-local}.
At each layer $i$ where acquaintance opportunities appear, consider the partition $\mathcal P_i$ whose parts are the pairs of qubits appearing in the acquaintance opportunities together with singleton parts for any unpaired qubits at the boundary. 
To obtain a $3$-complete linear swap network, we add $2$-swap networks corresponding to the partition $\mathcal P_i$, as shown in the bottom half of Figure~\ref{fig:3-local}.
The $3$-way acquaintance opportunities, where $3$-local gates (or compilations of them to $1$-and $2$-local gates) can be  added, are interspersed between the generalized swaps making up the $\mathcal P_i$-swap network, as shown in the top half of Figure~\ref{fig:4-local}. 
We make use of the property that for any two pairs of logical qubits involved in a 2-swap, each triple consisting of one of the pairs and one qubit from the other pair is mapped to three contiguous physical qubits either before or after the swap.
This ensures that overall every triple of logical qubits is acquainted because any triple $T$ is the union of a pair and a third qubit.
The 2-CCL network ensures that the pair is adjacent at some point, and thus a part $S$ of some partition $\mathcal P_i$.
The third qubit is necessarily in some other part $S'$ of the same partition, so that at some point in the $\mathcal P_i$-swap network there is a 2-swap network involving $S$ and $S'$, ensuring that the triple $T$ is acquainted.
(Actually, it is acquainted thrice, because there are three pairs $S$ for which the preceding logic applies.)
There are exactly $n$ $2$-swap networks inserted, and each $2$-swap gate can be implemented in depth $3$ using standard swap gates, for a total depth of approximately $(3/2) n^2 (\tau_2 + \tau_3) = \Theta(n^2)$.

\subsection{General $k$-qubit gates}

\begin{figure*}
\includegraphics[width=\textwidth]{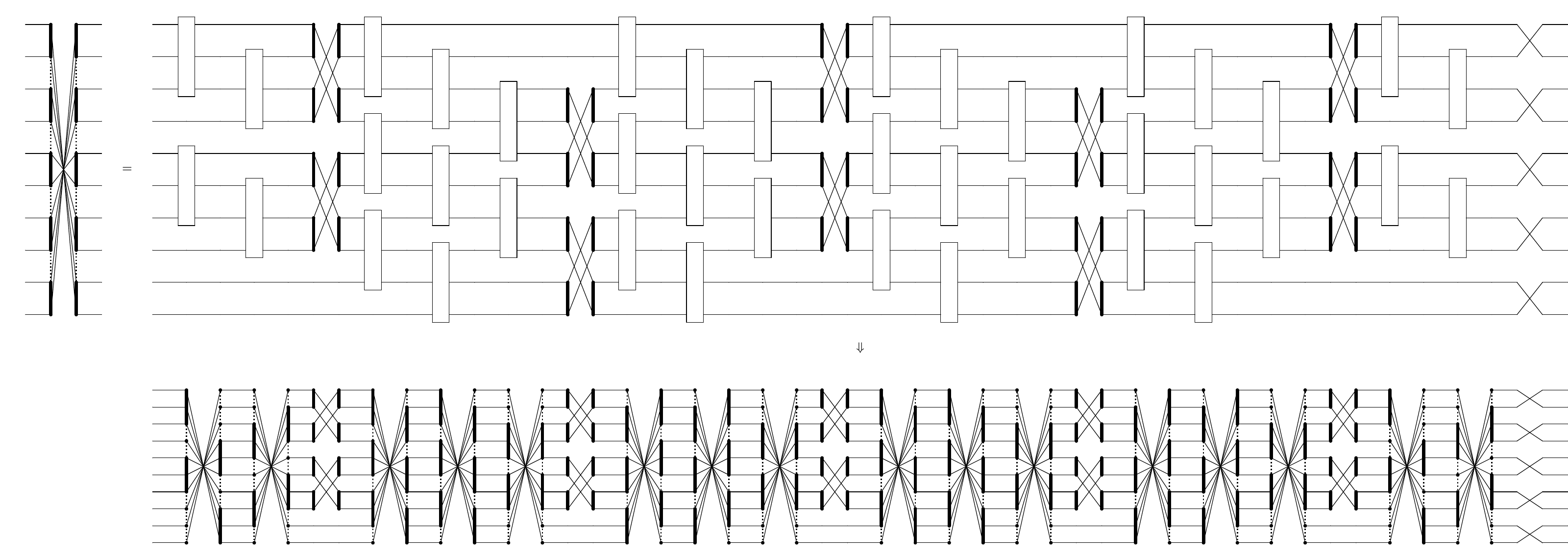}
\caption{\label{fig:4-local}
Top: Notation and decomposition for complete 2-swap network for acquainting each pair of qubits in the initial partition with every other third qubit.
Note how the $3$-qubit acquaintance opportunities are almost perfectly parallelized; this helps significantly when recursing further.
Bottom: Swap network for acquainting every set of $4$ qubits such that $2$ of the $4$ qubits were paired in the partition of the originating $2$-swap network,
formed by replacing each layer of acquaintance opportunities in the complete $2$-swap network with a complete $3$-swap network, in the same manner as in Figure~\ref{fig:3-local}.
In the swap network to acquaint \emph{every} set of $4$ qubits, this replacement (i.e., from top to bottom of this figure) is done for every complete $2$-swap network in the circuit for acquainting every set of $3$ qubits, i.e., for every other layer in the bottom of Figure~\ref{fig:3-local}.
}
\end{figure*}

The above ideas generalize to arbitrary $k$.
The construction is recursive.
First, construct the network to implement all $(k-1)$-qubit gates.
Then replace every layer $i$ of acquaintance opportunities with the corresponding $\mathcal P_i$-complete swap network, inserting $1$-swaps and acquaintance opportunities between the layers of $(k-1)$-swaps in order to acquaint each set of $k-1$ qubits with each qubit in the other set of $k-1$ qubits with which it will be swapped.
Specifically, when inserting a $(k-1)$-swap involving two sets of $(k-1)$ qubits each, we want to ensure that each set of $k$ qubits consisting of one of the sets and one qubit from the other set is mapped to $k$ contiguous physical qubits either before or after the swap.
For $k=2$, this is the case without additional swaps.
For larger $k$, this can be achieved by adding swaps that bring half each of set to the ``interface'' between them before the swap (the half closest to the interface), and the other half to the interface afterwards (when it will then be the closer half). 
This ensures that overall every set of $k$ logical qubits is acquainted because any such set $T$ is the union of a set $S$ of $k-1$ qubits and a $k$th qubit $t$.
Suppose we start with a swap network that acquaints every set of $k-1$ qubits, and in particular $S$, so that $S$ is a part of some partition $\mathcal P_i$ (corresponding to acquaintance layer $i$ in the starting swap network) in the recursive step.
The $k$th qubit $t$ is necessarily in some other part $S'$ of the same partition $\mathcal P_i$, so that at some point in the $\mathcal P_i$-swap network there is a $(k-1)$-swap involving $S$ and $S'$, ensuring that the set $T$ is acquainted.
(Actually, it is acquainted at least $k$ times, because there are $k$ sets $S \subset T$ for which the preceding logic applies.)

Each $(k-1)$-swap network has depth at most $n$ in terms of $(k-1)$-swap gates.
A $k$-swap gate has depth at most $2k-1$ in terms of standard swap gates, and the additional swaps for bringing inner qubits to the interface add depth $k-2$ at each swap.
Therefore, if we have a depth $O(n^{k-2})$ construction for all $(k-1)$-qubit gates, we can use that to get an $O(n^{k-1})$ depth construction for all $k$-qubit gates.
The base case is the linear-depth 2-CCL swap network for $2$-qubit gates.
Figures~\ref{fig:3-local}~and~\ref{fig:4-local} show the steps for $k=4$.
Lower-locality gates can be included in one of two ways, or a combination thereof.
First, they can be incorporated directly into the highest-locality gates.
Alternatively, the lower-locality acquaintance opportunities can be kept when recursing.

Using this recursive method yields a significant amount of redundancy with respect to the number of times that each set of $k$ qubits can be acquainted.
For applications in which the gates do not commute, this can be exploited in two ways.
First, distributing the gates over all possible acquaintance opportunities may lead to smaller Trotter errors.
Second, for each gate a possible acquaintance opportunity may be chosen randomly.
In other words, the swap network can be considered as a family of swap networks, each corresponding to a particular Trotter order; prior work shows that such random Trotter orderings may be helpful~\cite{childs2018faster}.

\subsection{Alternative for $3$-local}
Here we present an alternative construction for sets of $3$-local gates.
Its depth is similar to that of the other given, but it doesn't obviously generalize. We include it for two reasons:
it demonstrates a potentially useful property of complete linear swap networks,
and it may be better when applied to specific hardware devices.

Note that in the 2-complete swap network, every pair of logical qubits that is initially at distance $2$ from each other remains so, except near the ends of the line.
Furthermore, every other logical qubit passes through them at some point.
For our purposes, this means that in the course of the 2-complete swap network we can execute any $3$-local gate such that some pair of the three logical qubits on which it acts is at distance $2$ at the start of the network.

Consider a sequence of mappings labeled by $\Delta = 1, \ldots, n/2$.
In the mapping labeled by $\Delta$, the logical qubits 
$\left(1, 1 + \Delta, 1 + 2\Delta, \ldots, 2, 2 + \Delta, 2 + 2\Delta, \ldots, \Delta, 2 \Delta, 3\Delta, \ldots, \right)$ are mapped to physical qubits 
$\left(1, n, 3, n-1, 5, \ldots, \lfloor n /2 \rfloor + 1\right)$, respectively.
Any triple of logical qubits contains at least one pair that are mapped to physical qubits at distance $2$ in at least one of the $n/2$ mappings.
The construction is thus: alternate between 1) 2-complete swap networks with initial assignments given by the mappings, and 2) sorting networks to get to the next mapping. 
The 2-complete swap networks have depth $n$ and the sorting networks depth at most $n$, so overall the total depth is at most $2n \cdot (n/2) = n^2$.

\section{Unitary Coupled Cluster}\label{sec:unitary-coupled-cluster}

\begin{figure*}
\includegraphics[width=\textwidth]{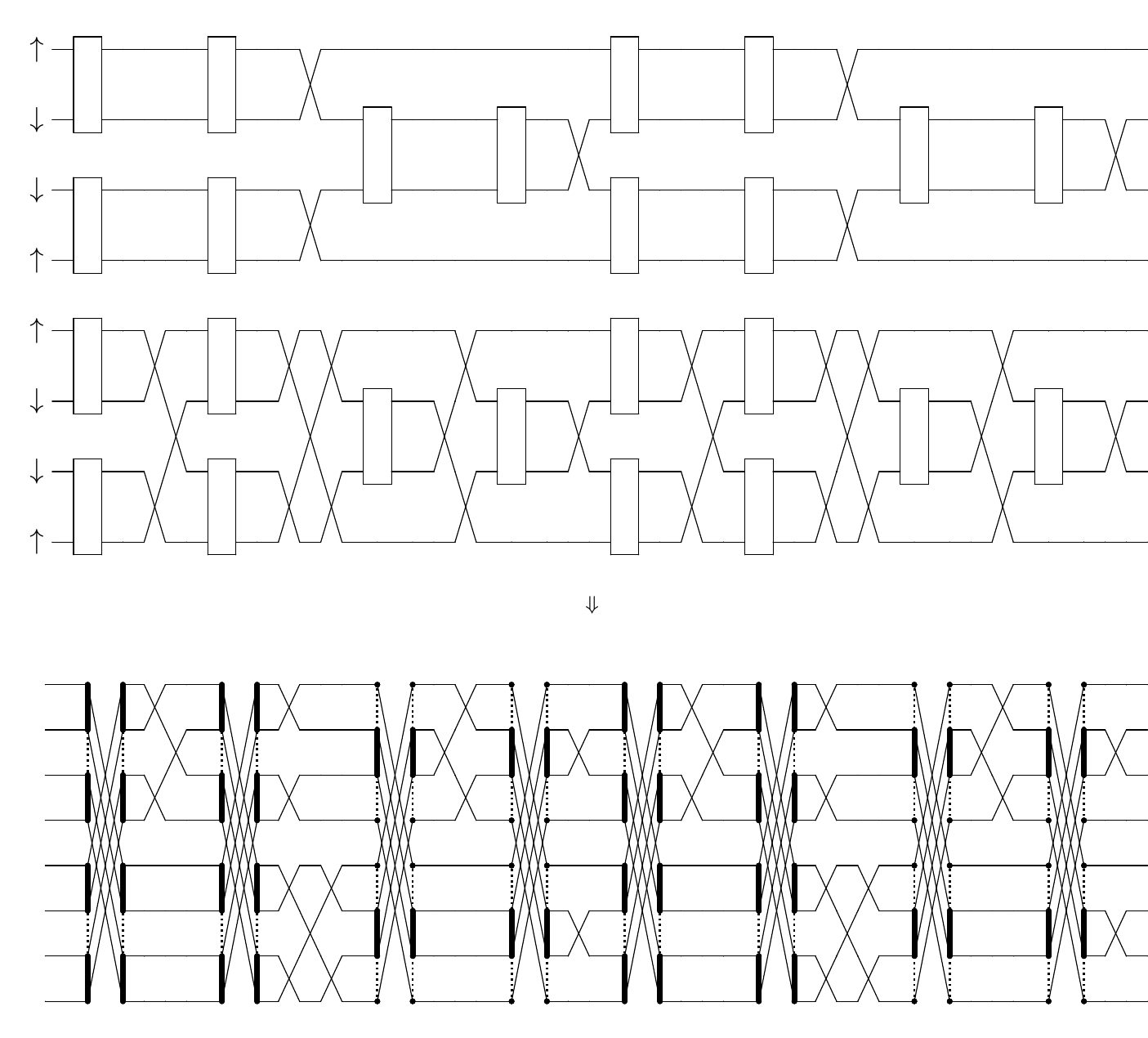}
\caption{\label{fig:UCCSD}
Construction of the swap network for a UCCSD operator (Equation~\ref{eq:UCCSD-operator}) with two occupied and two virtual spatial orbitals.
On top is an intermediate form of the construction useful for reasoning through the logic of its structure;
on bottom is the actual network.
On the first four qubits of the top half is a complete swap network with the acquaintance opportunity layers repeated twice.
On the second four qubits are four concatenated complete swap networks, one for each acquaintance layer (before repetition) of the complete swap network on the first four qubits.
The spins of the orbitals in the initial mapping to qubits is indicated;
with this initial mapping, the parity of the spins of the orbitals to be acquainted in each layer of the complete swap networks is the same (either all $\uparrow \uparrow$ or $\downarrow \downarrow$, or all $\uparrow \downarrow$).
For every pair of occupied spin orbitals with some parity and every pair of virtual spin orbitals with the same parity, there is some layer of the combined swap network in which both pairs are simultaneously (but separately) acquainted.
The construction is completed by replacing each acquaintance layer with a bipartite swap network over the occupied and virtual orbitals, which then acquaints the union of every such pair of pairs of spin orbitals.
(An example bipartite swap network is shown in Figure~\ref{fig:bipartite-swap-network}.)
}
\end{figure*}

In this section, we describe how the techniques of this paper can be used to implement Trotterized versions of three different types of unitary coupled cluster ansatz with a depth scaling that is optimal up to constant prefactors. 
We present the details for the standard unitary coupled cluster method with single and double excitations from occupied to virtual orbitals (UCCSD)~\cite{peruzzo2014variational,mcclean2016theory, romero2018strategies}, a unitary coupled cluster that includes additional, generalized, excitations (UCCGSD)~\cite{nooijen2000can,wecker2015progress}, and a recently introduced ansatz that is a sparsified version of UCCGSD (k-UpCCGSD)~\cite{lee2018generalized}.

The standard unitary coupled cluster singles and doubles ansatz is given by
\begin{equation}\label{eq:unitary_coupled_cluster_2}
    |\psi\rangle = e^{T - T^\dagger}|\phi_0\rangle,
\end{equation}
where \(|\phi_0\rangle\) is the Hartree-Fock state, \(T = T_1 + T_2\),
\begin{equation}\label{eq:UCCSD-operator}
\begin{split}
T_1 &= \sum_{\substack{i \in \mathrm{occ} \\ a \in \mathrm{vir}}} t_i^a a_a^{\dagger} a_i, \\
T_2 &= \sum_{\substack{i, j \in \mathrm{occ} \\ a, b \in \mathrm{vir}}} t_{i,j}^{a,b}
    a_a^{\dagger} a_b^{\dagger} a_j a_i.
\end{split}
\end{equation}

The $i$ and $j$ indices range over the $\eta$ ``occupied'' orbitals (those which are occupied in the Hartree-Fock state \(|\phi_0\rangle\)) and the $a$ and $b$ indices over the $n - \eta$ ``virtual'' orbitals (those which are unoccupied in \(|\phi_0\rangle\)).
A Trotter step of the corresponding unitary has $\binom{\eta}{2} \binom{n - \eta}{2}$ $4$-local gates.

These can be implemented in $O(\eta n^2)$ depth, as shown in Figure~\ref{fig:UCCSD}.
First, we assign the occupied orbitals to the first $\eta$ physical qubits $(1, \ldots, \eta)$ and the virtual orbitals to the last $n-\eta$ physical qubits $(\eta + 1, \ldots, n)$.
We have a 2-complete swap network on the occupied orbitals.
In between every swap layer thereof, we do a 2-complete swap network on the virtual orbitals.
For every pair of occupied orbitals and every pair of virtual orbitals, there is a layer in this composite network such that the pairs are simultaneously adjacent.
Thus, if we then insert a final 2-swap network with appropriate partitions at every layer, then every set of $2$ occupied orbitals and $2$ virtual orbitals will be adjacent at some point and a $4$-local gate can be implemented on them.
There swap depth of just the 2-complete swap networks is 
$\eta(n-\eta + 1) \tau_2 = \Theta(\eta n)$.
Before each one, a $2$-swap network is inserted with an average depth of $(n + 2)(3\tau_2 + \tau_4) = \Theta(n)$.
Overall, this yields the claimed $\Theta(\eta n^2)$ depth.
The coefficient of the leading term in the depth can be halved by accounting for the fact that we are typically interested in implementing only those excitations that are spin-preserving.
If we initially order the spin orbitals within the sets of occupied and virtual orbitals by $\uparrow, \downarrow, \downarrow, \uparrow, \uparrow, \ldots$, then the parity of the spins of the pairs of orbitals acquainted in each layer of the \(1\)-swap networks alternates, and we only need to do a bipartite \(2\)-swap network when the spin parities of the layers of the two sets coincide.

A more general version of the unitary coupled cluster ansatz is obtained by allowing excitations between any pair of orbitals. Rather than the cluster operators given in Equation~\ref{eq:UCCSD-operator}, we use 
\begin{equation}\label{eq:UCCGSD-operator}
\begin{split}
T_1 &= \sum_{p, q} t_p^q a_q^{\dagger} a_p, \\
T_2 &= \sum_{p, q, r, s} t_{p,q}^{r,s}
    a_r^{\dagger} a_s^{\dagger} a_q a_p,
\end{split}
\end{equation}
where the indices $p$, $q$, $r$, and $s$ are allowed to range over the entire set of orbitals (except that we often disallow excitations that do not preserve spin). 
It has been shown that the inclusion of these ``generalized' singles and doubles greatly increases the ability of unitary coupled cluster to target the kind of strongly correlated states that pose the greatest challenge for quantum chemical calculations on a classical computer~\cite{wecker2015progress, lee2018generalized}. 
A Trotter step for unitary coupled cluster with generalized singles and doubles may be implemented by a straightforward application of the techniques for implementing 4-local gates described in Figure~\ref{fig:4-local}. 
That construction also yields the optimal scaling here, enabling the execution of all \(\Theta(n^4)\) gates operations corresponding to the terms in Equation~\ref{eq:UCCGSD-operator} using a circuit of depth \(\Theta(n^3)\).
One possibility for exploiting spin symmetry is as follows.
Start with an initial mapping in which the orbitals of one spin are mapped to the first half of the physical qubits and those of the other spin to the second half.
Then apply the quartic swap network to each half of the qubits in parallel, thus acquainting all sets of four orbitals with the same spin.
Then apply a double bipartite swap network, of the sort used for UCCSD, to acquaint every set of four orbitals such that there are two orbitals of each spin.

As a final example of the utility of a swap network approach to circuit compilation, we describe the implementation of a sparse version of the unitary coupled cluster operator with generalized singles and doubles recently developed by Lee at al.~\cite{lee2018generalized}.
Rather than the full set of double excitations as in Equation~\ref{eq:UCCGSD-operator}, this variant of unitary coupled cluster uses only those double excitations that transfer two electrons with opposite spins from one spatial orbital to another.
The resulting cluster operators,
\begin{equation}\label{eq:UpCCGSD-operator}
\begin{split}
T_1 &= \sum_{p, q, \alpha} t_p^q a_{q\alpha}^{\dagger} a_{p\alpha}, \\
T_2 &= \sum_{p, q} t_{p,p}^{q,q},
    a_{q \uparrow}^{\dagger} a_{q \downarrow}^{\dagger} a_{p \downarrow} a_{p \uparrow}
\end{split}
\end{equation}
contain only \(\Theta(n^2)\) terms and can be implemented in \(\Theta(n)\) depth using the approach detailed below.

Recall our prior observation that, throughout the execution of a complete $1$-swap network, every pair of logical qubits that is initially at distance 2 from each other will remain so. 
Furthermore, every such pair of logical qubits will become adjacent to every other pair. 
Therefore we begin by ordering the fermionic modes \((1\uparrow, 2\uparrow, 1\downarrow, 2\downarrow, 3\uparrow, 4\uparrow, 3\downarrow, 4\downarrow, \ldots)\).
Then, by executing a 2-complete swap network, we bring the fermionic modes involved in each of the 2-local and 4-local terms in Equation~\ref{eq:UpCCGSD-operator} adjacent to each other at some point. 
We show an example for \(n=8\) in Figure~\ref{fig:UpCCGSD} below. 

\begin{figure*}
\includegraphics[width=\textwidth]{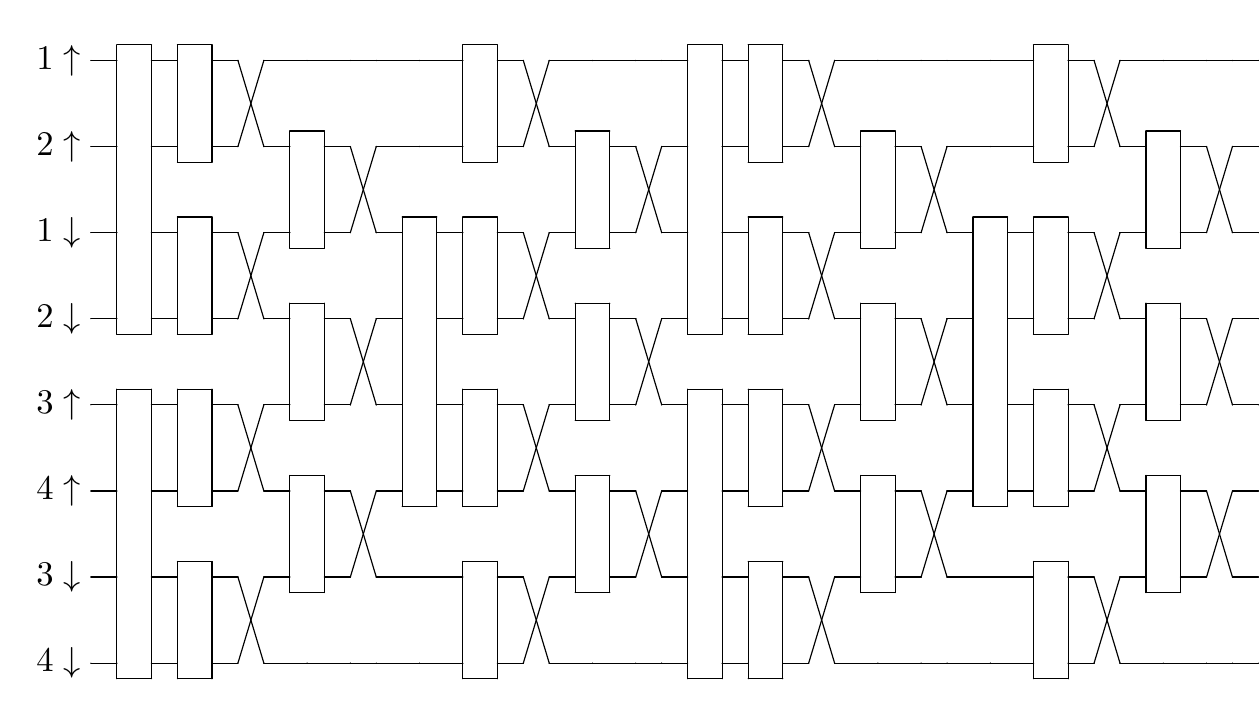}
\caption{\label{fig:UpCCGSD}
The swap network for a UpCCGSD operator (Equation~\ref{eq:UpCCGSD-operator}) with four spatial orbitals.
The initial assignment of spin orbitals to qubits is indicated; the important feature is that the two spin orbitals for each spatial orbital are assigned to qubits at distance $2$ from each other.
They then \emph{stay} at distance $2$ from each other throughout the evolution of the swap network (except temporarily at the edges).
The swaps are exactly the same as in the standard $1$-swap network, except that a layer of $4$-local acquaintance opportunities is inserted before every other swap layer, allowing the four spin orbitals corresponding to a pair of spatial orbitals to be acquainted. 
}
\end{figure*}

\section{Conclusion}\label{sec:conclusion}
We have introduced and instantiated an instance-independent approach to quantum circuit routing.
This instance-independent approach has a distinct advantage among the growing number of alternatives for addressing the limited connectivity of physical devices: it requires effectively no marginal classical computation per instance.
Of course, there is the corresponding disadvantage that it cannot in general achieve instance-specific optimality.
However, for many applications, including the fermionic simulation tasks we addressed, all instances of a given size share a topology.
For applications in which this is not the case, instance-independent swap networks can nevertheless provide a starting point for further optimizations.
(Regardless, simple local optimizations, such as removing any two swap gates in a row on the same pair of qubits, should be used to tighten up the swap networks presented here in any practical implementation.)

Another limitation of our approach is the complete separation of the decomposition aspect of compilation from the routing aspect; perhaps a better compilation can be found by solving these together at once. 
Nevertheless, given the hardness of the compilation problem in its full generality, we expect that this separation will in general be useful in balancing quality of the solution found with the time to find it. 

We have made a connection between the quantum circuit routing problem and the minor embedding problem in quantum annealing. 
This analogy should not be taken too mathematically, especially when considering the ordered variant of the routing problem, but may still be of value in encouraging the lifting of ideas from the significant body of theoretical and applied work on minor embedding.
For example, the separation of gate decomposition and circuit routing can be thought of as corresponding to the separation of the parameter-setting and minor-embedding aspects of compilation in quantum annealing. 

While our motivation and focus is NISQ-era devices, our results may continue to be applicable even with full error correction.
In the surface code, for example, the dominant cost with respect to both time and qubits is the implementation of T gates; in comparison, the cost of swap gates are negligible, and thus so is the overhead in overcoming limited connectivity.
However, even error-corrected devices benefit from parallelization.
Our constructions for swap networks imply a scheme for parallelization, which may be of use independent of any mapping to physical qubits.
For problems arising from the Jordan-Wigner transformation of fermionic Hamiltonians, the swap networks are just as useful even with ``free'' (fermionic) swap gates.
In that case, the locality to be addressed is not spatial locality but the number of qubits that each gate acts on, which must be bounded even in the error-corrected regime.
The same applies to proposed ion trap implementations with effectively all-to-all connectivity.

There are several directions for future work.
Of most practical interest is lowering the abstraction level.
That is, using the high-level constructions presented here to compile specific families of circuits to low-level hardware with restricted gate sets and variable durations.
This is a necessary step in a more general program of directly comparing swap network-based methods to alternative approaches, with respect to quantum resources, basis-set errors, Trotter errors, etc.
Furthermore, for some algorithms, there is freedom in the choice of operator at certain stages in the algorithm. For example, for the alternative mixers in the Quantum Alternating Operator Ansatz~\cite{Hadfield17_QAOAGen}, while reordering the gates gives different mixers, it is an open research question as to which mixers and which orders provide the best performance. Given the limited depth of NISQ devices, the efficiency with which qubit routing can be achieved for a given operator significantly impacts the choice of operator. The techniques described here provide a key step toward exploring these trade-offs.

There is also further work to be done in the present abstraction level.
Specifically, our construction for $4$-local gate sets is likely suboptimal with respect to constant factors, and may be improved. 
The same goes for $k > 4$.
We also focus only on the routing problem for unordered sets of gates, in which there is no precedence structure to be enforced on the logical gates; examples of solutions to the ordered problem would significantly broaden the usefulness of this approach.  
One limited example would be the iterated circuits of a complete variational algorithm or Trotter-based simulation, whereas in the present work we focused on a single iteration.

More generally, with this work we have established a foundation for designing swap networks for more applications and more architectures.
A more comprehensive understanding of how well different architectures support the topologies of different applications can be the foundation for co-design in both directions: in one direction motivating new architectures by how well they are suited generally or specifically to applications, and in the other direction tweaking problems in a way that doesn't degrade the value of their solution but that makes them more efficiently solvable on a quantum computer.

\begin{acknowledgements}
This work was supported by the U.S. Department of Energy, Office of Science, 
Office of Advanced Scientific Computing Research, 
Quantum Algorithm Teams Program, under contract number DE-AC02-05CH11231. We are also grateful for support from the NASA Ames Research Center, the NASA Advanced Exploration systems (AES) program, the NASA Transformative Aeronautic Concepts Program (TACP), and from the AFRL Information Directorate under grant F4HBKC4162G001.
B.O. was supported by a NASA Space Technology Research Fellowship.
\end{acknowledgements}

\appendix
\section{Instance-independent embedding for quantum annealing}
\label{sec:quantum-annealing}
Quantum annealing is an alternative model of quantum computation for minimizing a classical pseudo-Boolean function $f: {\{\pm 1\}}^n \to \mathbb R$, in which the Hamiltonian is slowly changed from an initial Hamiltonian $H_{\mathrm{init}}$ into the problem Hamiltonian $H_f$, whose ground state(s) we would like to find. 
Often, the desired Hamiltonian $H_f$ cannot be implemented directly on a physical quantum annealer due to limited connectivity.
To overcome this limitation, each logical qubit in $H_f$ can be mapped to a connected set of physical qubits which are coupled together with a ferromagnetic field that induces them to take on the same value.
In the standard case in which $H_f$ is 2-local (i.e., $f$ is quadratic), it can be considered as a graph, and this mapping from logical to physical qubits as a minor embedding into the hardware graph.
For example, Choi~\cite{choi2011minor-embedding} gave a family of minor embeddings of the complete graph into a so-called Triad hardware graph (similar to the Chimera hardware graph used by D-Wave) in which the number of physical qubits scales quadratically with the number of logical qubits, which is optimal for bounded-degree hardware graphs. Zaribafiyan et al.~\cite{zaribafiyan2017systematic} provide a deterministic embedding for Cartesian product graphs. 

In practice, problem graphs of interest are usually much sparser than the complete graph, or the Cartesian product graphs, and so using an embedding for the complete graph is likely to use more physical qubits than necessary.
Specifically, most problems run on the D-Wave quantum annealer make use of D-Wave's heuristic embedding software~\cite{cai2014practical}. 
Many practitioners thus use instance-specific embeddings to maximize the use of scarce resources.
The problem, however, is the difficulty of finding such instance-specific embeddings.
An approach similar to the one we used for quantum circuits can be taken.
Instead of using an embedding of either the complete graph (which is trivial to find but resource-inefficient) or a single problem graph (which is harder to find but more resource-efficient), one can use an embedding of a ``supergraph'' of a class of problem graphs.
Such an embedding can be found either manually or algorithmically, but in any case can be reused for any instance in the class with negligible marginal cost.
This approach thus strikes a potentially valuable balance between the two existing ones.

\section{Lower bounds}
\label{sec:lower-bounds}

The optimality of the complete swap network is easy to show. 
$\binom{n}{2}$ logical gates are executed in $n$ almost perfectly parallelized layers.
In a reasonable accounting in which any $2$-qubit gate on adjacent qubits can be done in unit time, the logical qubits and swaps can be combined into one.
However, for more complicated cases the reasoning becomes more involved. This section gives some methods for lower bounding the depth of solutions to the (unordered) circuit embedding problem.
In particular, the lower bounds are on the depth of the $2$-qubit swaps only, i.e., the ``swap depth''.
For a bounded-degree physical graph and bounded-locality logical graph, the logical gates that can be executed with a single, fixed mapping of logical to physical qubits, i.e., that after an swap layer, can be executed in $O(1)$ depth.
In such cases, which comprise almost all of practical interest, exact lower bounds on the swap depth thus yield scaling lower bounds on the total depth.

\subsection{Acquaintance time}\label{subsec:acquaintance}

Benjamini et al.\ defined~\cite{benjamini2014acquaintance} the \emph{acquaintance time} of a graph $G$, denoted $\mathcal{AC}(G)$ as follows.
Consider placing an agent at each vertex of the graph and a series of matchings~\footnote{A matching is a set of mutually disjoint edges of a graph.} of the graph.
Each matching corresponds to simultaneously swapping the agents on the vertices of each edge.
Such a a sequence of matchings of $G$ is a \emph{strategy for acquaintance} in $G$ if every pair of agents are adjacent in the graph $G$ at least once.
The acquaintance time is the number of rounds (matchings) in the shortest strategy for acquaintance (and is finite if and only if the graph is connected).

This notion of strategies for acquaintance is a useful if limited abstraction for compiling quantum circuits around geometric constraints.
As is, a strategy for acquaintance corresponds to a compilation of all $2$-local gates in a hardware graph $G$, with agents corresponding to logical qubits, vertices corresponding to physical qubits, and edges of matchings to swap gates.
A gate between two logical qubits can be implemented at any point that that they can become ``acquainted''.
This level of abstraction has the advantage and disadvantage that it disregards the exact nature of the gates.
This makes it extremely general but also constructions within it somewhat approximate.
For example, in a strategy for acquaintance, it is permissible for an agent to become acquainted with more than one other agent in a single round, while the corresponding $2$-local gates would need to be implemented sequentially.

Nevertheless, known results about acquaintance times~\cite{benjamini2014acquaintance,angel2016tight} can be interpreted in the context of quantum circuit embedding.
For example, that the acquaintance time of the path graph $P_n$ is $n-2$ provides an alternative proof of the optimality of the complete linear $1$-swap network. 
Interestingly, the acquaintance time of the barbell graph $B_n$ (two fully connected halves connected by a single edge) is also $n-2$.
Generally, it is known that for a graph $G$ of maximum degree $\Delta$, $\mathcal{AC}(G) = \min\{O(n^2 / \Delta), 20 \Delta n\}$, which in particular implies that for any graph $\mathcal AC(G) = O(n^{3/2})$. 
There are also hardness results: $\mathcal{AC}(G)$ is NP-hard to approximate within a multiplicative factor of $2$ or within any additive constant factor.

A strategy for acquaintance as defined above requires that \emph{every} pair of agents become acquainted.
However, it will often be the case that we care only about certain pairs of agents, or larger-sized sets of agents.
We now define a generalization of acquaintance time that may be of value in finding lower bounds in such cases.
Let $H$ be the hypergraph whose vertices correspond to the agents and whose hyperedges correspond to the sets of agents that we would like to acquaint.
We can then define a \emph{strategy for $H$-acquaintance in $G$} as an initial (injective) mapping $\sigma$ of the vertices of $H$ to the vertices of $G$ and a sequence of matchings as above such that, for every edge $\{i_1, \ldots, i_k\}$ of $H$, if agent $i_1$ is placed on vertex $\sigma(i_1)$ in $G$, $i_2$ on vertex $\sigma{(i_2)}$, and so on, then the set of agents $\{i_1, \ldots, i_k\}$ can be acquainted at some point.
Whether a set of agents can be acquainted given their locations on the vertices of $G$ can be specified in one of two ways.
In the first case, $G$ itself is a hypergraph and the agents can be acquainted if their positions $\{\sigma_t(i_1), \ldots, \sigma_t(i_2)\}$ are a hyperedge of $G$, where $\sigma_t(i)$ is the location of agent $i$ after $t$ rounds.
In the alternative, $G$ is a simple graph, and the agents can be acquainted if their positions form a connected subgraph of $G$.
The latter is closer to our application of strategies for acquaintance: the physical graph $G$ specifies on which pairs of qubits a $2$-qubit gate can be applied, and higher-locality gates are decomposed using such $2$-qubit gates.
The \emph{$H$-acquaintance time} of $G$, denoted $\mathcal{AC}_H(G)$ then is the minimal size of a strategy for $H$-acquaintance in $G$.
Note that this definition does not assume that $|V(H)|=|V(G)|$.

\subsection{Circuit embeddings as minor embeddings}
\begin{figure}
\includegraphics[width=\columnwidth]{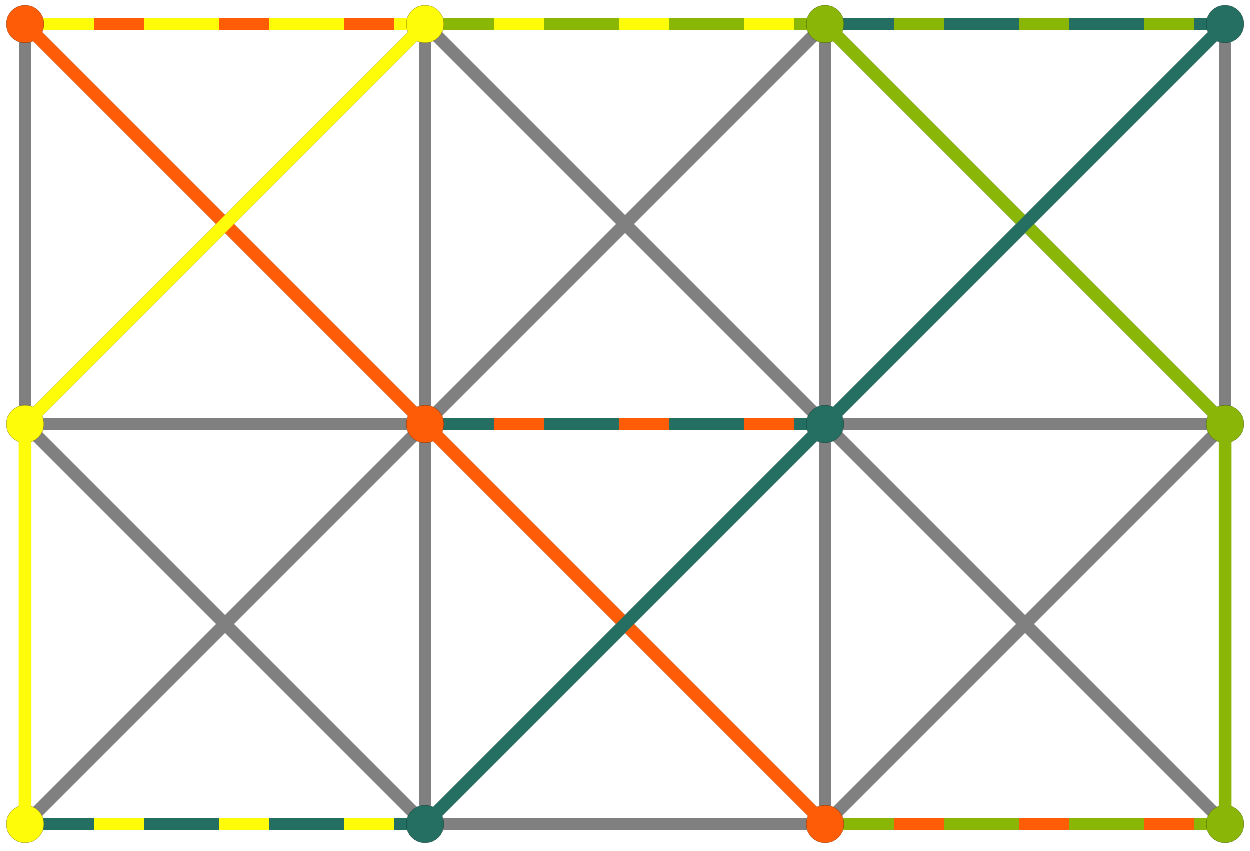}
\caption{\label{fig:strategy-minor-embedding}
A $2$-round strategy for $K_4$-acquaintance in $P_4$ as a minor embedding of $K_4$ into $P_4 \boxtimes P_3$.
Each set of solid points and lines of a given color indicates a vertex model.
Each bi-colored dashed line indicates an edge model.
Solid gray lines indicate unused edges of $P_4 \boxtimes P_3$.
}
\end{figure}

This section assumes that the reader is familiar with the basic ideas of graph minor embeddings and treewidth; see Klymko et al.~\cite{klymko2014adiabatic} for a brief introduction to these ideas in a related context.
All graphs in this section will be assumed to have edges of size $2$.

Consider a strategy for $G$-acquaintance in $\Gamma$ with $d$ rounds.
Let $H = \Gamma \boxtimes P_{d + 1}$ be the strong product of $\Gamma$ and the path graph on $d+1$ vertices.
That is, 
\begin{align}
V(H) &= \left\{
    (v, t) \middle| 
    v \in V(\Gamma), t \in \left\{0, \ldots, d\right\}
\right\}, \\
E(H) &= \left\{
    \left\{(v, t), (v', t')\right\} \middle|
    v=v' \lor \left\{v, v'\right\} \in E(\Gamma),
    |t - t'| \leq 1
\right\}.
\end{align}
The strategy for $G$-acquaintance in $\Gamma$ can be interpreted as a graph minor embedding of $G$ into $H$ as follows.
Figure~\ref{fig:strategy-minor-embedding} shows an example for $G=K_4$ and $\Gamma=P_4$.
The ``agents'' are the vertices of $G$.
The vertex model of $v \in G$ is the set of vertices $\left\{(\sigma_t(v), t) | t \in \{0, \ldots, d\}\right\} \subset V(H)$  corresponding to the series of assignments of $v$ to vertices of $\Gamma$.
Note that this vertex model is connected (indeed, a simple path) and that the vertex models of distinct vertices are disjoint, by the properties of an acquaintance strategy.
The edge model of an edge $\{v, w\} \in E(G)$ is
$\left\{(\sigma_t(v), t), (\sigma_t(w), t)\right\} \in E(H)$ for some round $t$ in which the vertices $v$ and $w$ are assigned to adjacent vertices of $\Gamma$.
For any graphs $A$ and $B$, if $A$ is a minor of $B$, then $\mathrm{pw}(A) \leq \mathrm{pw}(B)$ and $\mathrm{tw}(A) \leq \mathrm{tw}(B)$, because any path or tree decomposition for $B$ can be converted into one for $A$ by edge-contracting the vertex models, without increasing the relevant width.
In our case, we have shown that $G$ is a minor of $H = \Gamma \boxtimes P_{d + 1}$ whenever there exists a $d$-round strategy for $G$-acquaintance in $\Gamma$.
Therefore,
\begin{equation}\label{eq:pw-lb-general}
\mathrm{pw}(G) \leq 
\mathrm{pw} \left(\Gamma \boxtimes P_{\mathcal{AC}_G(\Gamma) + 1}\right),
\end{equation}
and similarly for treewidth.

We show now that, for an arbitrary graph $G$ on $n$ vertices, the pathwidth $\mathrm{pw}(G)$ is at most about one more than the $G$-acquaintance time in the path graph $P_n$,
\begin{equation}\label{eq:pw-lb-path}
\mathrm{pw}(G) \leq 
2 \left\lceil \frac{\mathcal{AC}_G(P_n)}{2} \right\rceil + 1.
\end{equation}
We do so by explicitly constructing a path decomposition of a graph from a strategy for $G$-acquaintance in $P_n$.
Consider such a strategy and let $\sigma_t(v) \in P_n$ be the assignment of vertex $v\in G$ after round $t$.
We can construct a path decomposition with $n-1$ bags as follows.
Each bag corresponds to an edge of $P_n$ and contains all the vertices of $G$ that are assigned to an vertex of $P_n$ adjacent to $e$.
The bags form the path graph $P_{n-1}$ corresponding to the line graph of $P_n$.
Each bag can contain at most 
$2\left\lceil d / 2 \right\rceil + 2$ 
vertices, where $d$ is the number of rounds in the strategy for $G$-acquaintance.
Lastly, the number of rounds in the strategy is at least the minimum number of rounds $\mathcal{AC}_{G}(P_n)$ and the pathwidth of the graph is at most the width of this decomposition, yielding the desired inequality.

One application of this inequality is yet another lower bound on the swap depth of a complete swap network.
Equation~\ref{eq:pw-lb-path} and the fact that $\mathrm{pw}(K_n) = n - 1$ imply that
\begin{flalign}
&& \mathrm{pw}(K_n) &= 
n - 1 
\leq 2 \left\lceil \frac{\mathcal{AC}(P_n)}{2} \right\rceil + 1 \\
\Rightarrow &&
\frac{n}{2} - 1 &\leq \left\lceil \frac{\mathcal{AC}(P_n)}{2} \right\rceil \\
\Rightarrow && 
\mathcal{AC}(P_n) &\geq 
\begin{cases}
  n - 2, n \text{\ odd}, \\
  n - 3, n \text{\ even}.
\end{cases}
\end{flalign}

Note that Equation~\ref{eq:pw-lb-path} is not necessarily tight for arbitrary graphs.
For example, consider the star graph $S_k$ for large $k$.
It has pathwidth 1~\footnote{Consider the decomposition in which there is a bag for each leaf containing that leaf and the internal vertex.},
but the minimum swap circuit depth is $\Omega(k)$.
More generally, caterpillar graphs exemplify the looseness of the above bound for the same reason; the minimum depth of a swap circuit for any graph scales linearly with the degree of the graph.

\bibliography{routing} 

\end{document}